\documentclass[a4paper]{article}

\usepackage[utf8]{inputenc}
\usepackage[T1]{fontenc}

\usepackage[a4paper,margin=2.5cm,right=3cm]{geometry}
\usepackage{pdflscape}
\usepackage{lmodern}

\usepackage{mathtools,amsmath, amssymb}
\usepackage{tensor,mathrsfs,upgreek,stmaryrd}
\usepackage[normalem]{ulem}

\usepackage{cite, aas_macros}
\usepackage[sort&compress,numbers]{natbib}

\usepackage[dvipsnames]{xcolor}
\usepackage{url}
\usepackage[colorlinks=true, citecolor=Blue, linkcolor=Red]{hyperref}

\newcommand{\cref}[1]{\red{\ref[1]}}

\let\t\tensor
\let\p\partial

\newcommand{\redsigma}{\sigma}
\newcommand{\redtau}{\tau}

\newcommand{\ptcheck}[1]{\ptc{checked on #1}}

\newcommand{\reda}[1]{\ptcr{change or addition or rewording}\color{red}}

\newcommand{\red}[1]{{\color{red}#1}}

\newcounter{mnotecount}[section]
\renewcommand{\themnotecount}{\thesection.\arabic{mnotecount}}

\newcommand{\mnote}[1]{%
\protect{\stepcounter{mnotecount}}%
\textsuperscript{$\bullet$\themnotecount}%
\marginpar{\raggedright\tiny\em$\hspace{-1em}\bullet$\themnotecount: #1}%
}

\newcommand{\ptc}[1]{\mnote{{\bf ptc:}#1}}
\newcommand{\ptcr}[1]{{\color{red}\mnote{{\color{red}{\bf ptc:}#1} }}}

\newcommand{\g}{\gamma}
\newcommand{\m}{\mathfrak{m}}
\newcommand{\redmusquared}{\mu^2_{\mathrm{P}}}
\newcommand{\redmu}{\mu_{\mathrm{P}}}

\let\t\tensor
\let\p\partial

\def\R{\mathbb R}

\def\mcL{{\mathcal L}}
\def\m{\mathfrak m}

\def\redsigma{ \sigma}
\def\redtau{\tau}
\def\ladder{\mathfrak D}

\usepackage{textgreek}
\def\permeability{\text{\textmu}}
\def\permittivity{\text{\textepsilon}}

\usepackage{authblk}

\renewcommand{\ptcheck}[1]{}

\begin{document}
\title{Proca Fields in Step-index Optical Fibres\thanks{Preprint UWThPh 2023-6}}

\author[1,2]{F.~Steininger}
\affil[1]{\footnotesize University of Vienna, Faculty of Physics, Vienna Doctoral School in Physics, Boltzmanngasse 5, 1090 Vienna, Austria}
\affil[2]{\footnotesize University of Vienna, Faculty of Physics and Research platform TURIS, Boltzmanngasse 5, 1090 Vienna, Austria}

\date{\today}
\maketitle

\begin{abstract}
The topic of a photon mass has received justified attention over the years with strong theoretical reasoning for its existence.
We derive a full description of mode solutions in step-index optical fibres on the basis of Proca's equation as a model for massive photons.
The no-go theorem recently found in coaxial waveguides~\cite{CMS22} does not appear in the optical fibre case.
\end{abstract}


\section{Introduction}
\label{sec:Proca}

The current physical paradigm considers photons to be massless, backed by more than a century of continuously improving experiments~\cite{Goldhaber08} and an increasingly mature theoretical framework.
Nevertheless, some mechanisms have been described~\cite{AdelbergerDvaliGruzinov, BONETTI2017203} which suggest that a small non-vanishing mass might be unavoidable.

A minimal coupling approach to introducing a photon mass into the theory comes from modifying the vacuum Maxwell Lagrangian to include a mass term $\redmu > 0$ via
\begin{equation}
	\mcL =
 \Big(
 - \frac{1}{4} \t{ \eta }{^\alpha^\beta} \t{ \eta }{^\rho^\sigma} \t{ F }{_\alpha_\rho} \t{ F }{_\beta_\sigma} -
  \frac 12 \frac{\redmusquared c^2}{\hbar^2}  \t{ \eta }{^\alpha^\beta} A_\alpha A_\beta + j^\alpha A_\alpha
  \Big)
   \sqrt{|\det \eta|}
  \,,
\label{eq:MaxwellL}
\end{equation}
with
\begin{equation}
	F_{\alpha \beta} = \p_\alpha A_\beta - \p_\beta A_\alpha
	\,.
\end{equation}
The external current $j^\alpha$ is assumed to fulfill $\nabla_\alpha j^\alpha = 0$, where $\nabla_\alpha$ is the covariant derivative with respect to $\eta_{\alpha \beta} = \text{diag} ( -1 , +1 , +1, +1)$, taken to be the Minkowski metric. Going forward we set $\hbar = c = 1$.
The derived field equations
\begin{equation}
	\eta^{\alpha \beta} \p_\alpha F_{\beta \gamma} - \redmusquared A_\gamma = - \eta_{\gamma \alpha} j^\alpha
	\,,
	\label{eq:Proca}
\end{equation}
are well known as Proca's equation~\cite{Proca36}.

Current estimates by the Particle Data Group put an upper bound of $\redmu < 10^{-18} \;\text{eV}$ on the photon mass~\cite{ParticleDataGroup:2022pth}, which is accumulated from a number of laboratory experiments and astronomical observations~\cite{Goldhaber08}.
Upcoming high precision photon interferometers as proposed by~\cite{Hilweg17, Mieling22} might be suitable as a new step in setting more stringent bounds on the mass $\redmu$.

This serves as motivation to search for a formulation of Proca's theory in linear isotropic media and its application to optical fibres.
In the case of the standard Maxwell  theory with massless photons, the usual modification of the Lagrangian is given by
\begin{equation}
	\mcL =
 \Big(
 - \frac{1}{4 \permeability} \t{ \gamma }{^\alpha^\beta} \t{ \gamma }{^\rho^\sigma} \t{ F }{_\alpha_\rho} \t{ F }{_\beta_\sigma} + j^\alpha A_\alpha
  \Big)
   \sqrt{|\det \eta|}
  \,,
\label{eq:MaxwelldielectricL}
\end{equation}
with Gordon's optical metric
\begin{equation}
	\t\gamma{^\alpha^\beta}
		= \t\eta{^\alpha^\beta}
		+ (1 - n^2) \t u{^\alpha} \t u{^\beta}\,,
\end{equation}
where the refractive index is defined as $n = \sqrt{ \permittivity \permeability}$, with permittivity $\permittivity$ and permeability $\permeability$ of the medium~\cite{Gordon23}.
The refractive index is taken to be a function of the spacetime coordinates $n = n (x)$.
Adding a mass term now comes with the ambiguity of choosing either $\gamma^{\alpha \beta}$ or $\eta^{\alpha \beta}$ as the tensor field appearing in the term quadratic in $A_\mu$ in \eqref{eq:MaxwellL}.
We remain agnostic of the choice for now, writing $\m \in \{\gamma, \eta\}$ and 
\begin{equation}
	\mcL =
 \Big(
 - \frac{1}{4 \permeability} \t{ \g }{^\alpha^\beta} \t{ \g }{^\rho^\sigma} \t{ F }{_\alpha_\rho} \t{ F }{_\beta_\sigma} -
  \frac 12 \redmusquared  \t{ \m }{^\alpha^\beta} A_\alpha A_\beta
  + j^\alpha A_\alpha
  \Big)
   \sqrt{|\det \eta|}
  \,.
\label{eq:L}
\end{equation}

We have shown in~\cite{CMS22} that the Proca equation does not allow for TEM modes in coaxial cables, no matter how small the mass $\redmu$ of the Proca photons is. Keeping in mind that the TEM modes are the main propagation modes in such cables, the analysis in~\cite{CMS22} shows that \emph{either} Proca photons do not exist, \emph{or} the idealised conductor model used there  is unrealistic. Therefore it is of interest to enquire whether a similar no-go result arises with Proca fields in step-index optical fibers. We show here that this is not the case: consistent solutions of the Proca equations can be found which are small perturbations of the standard Maxwell modes for all known modes. This is the case for both choices $\eta^{\alpha\beta}$ and $\gamma^{\alpha\beta}$ of the mass tensor $ \t{ \m }{^\alpha^\beta}$.

 It turns out the analysis is considerably more complicated with the choice $  \m =\eta$. So, from the point of view of simplicity of the theory, our analysis strongly points to the choice $ \m  =\gamma$, but does not eliminate the alternative.

Going forward, we introduce $G^{\alpha \beta} : = \gamma^{\alpha \rho} \gamma^{\beta \sigma} F_{\rho \sigma}$ as a macroscopic electromagnetic tensor and derive the field equations
\begin{equation}
	\sqrt{|\det \eta |}^{-1} \p _\alpha
 \big( \permeability^{-1} \sqrt{|\det \eta |} \,   \t{ G }{^\alpha^\beta}
 \big)
 - \redmusquared
   \t{ \m }{^\alpha^\beta} A_\alpha = -  j^\beta\,.
\label{eq:fieldeq}
\end{equation}

By taking the divergence of \eqref{eq:fieldeq} we find that the mass term implies the constraint
\begin{equation}
	\p _\alpha \left(\sqrt{|\det \eta |} \,
 \t{ \m }{^\alpha^\beta} A_\beta \right) = 0
\,,
\label{eq:constraint}
\end{equation}
which is similar to the Lorenz-gauge condition in the massless case.
Assuming the medium to be additionally homogeneous and non-magnetic, the field equations are further equivalent to a wave equation for the potential $A_\alpha$
\begin{equation}
	\Box_\g A_\beta - \t{ \g }{^\sigma^\rho} \nabla _\beta \nabla _\sigma A_\rho - \redmusquared A_\sigma \t{ \m }{^\sigma^\rho} \t{ \g }{_\rho_\beta} = -  \gamma_{\beta\gamma} j^\gamma
\,,
\label{eq:wave}
\end{equation}
where $\Box_\g := \g^{\alpha \beta} \nabla_\alpha \nabla_\beta$.
This also provides the equation for piecewise homogeneous media by matching the potential through discontinuities via interface conditions (see~\cite{Love83,Jackson98}).

\section{Step-index optical fibre}
\label{sec:Fibre}

We set out to solve the macroscopic Proca equations in an optical fibre.
As a reminder, an optical fibre consists of a cylinder of dielectric media with refractive index $n_1$ and radius $a$, surrounded by a layer of dielectric material with refractive index $n_2$.
The electromagnetic field propagates throughout both media and along the cylinder axis.
It is suggestive to use cylindrical coordinates $( t , r , \theta , z)$ in our description and call the inner region ( $r < a$ ) the core and the outer region ( $r > a$ ) the cladding.

Instead of trying to solve equation \eqref{eq:fieldeq} directly, it is preferable to search for solutions of the wave equation \eqref{eq:wave} independently in the core and in the cladding and match the solutions across the interface.

\subsection{Interface conditions}
\label{sec:interface}

The following equations have to be satisfied everywhere and thus in particular at the interface:
\begin{eqnarray}
	\sqrt{|\det \eta |}^{-1} \p _\alpha
 \big( \sqrt{|\det \eta |} \,   \t{ G }{^\alpha^\beta}
 \big)
 - \redmusquared
   \t{ \m }{^\alpha^\beta} A_\alpha
	&=&
	0
\,,
\label{7VI22.1a}
\\
	\p_{[\alpha} \t{ F }{_\beta_{\gamma]}}
	&=&
	0
\,,
\label{7VI22.1b}
\\
	\p_{\alpha} \left(  \sqrt {  | \det \eta | } \t{ \m }{^\alpha^\beta}  A_\beta \right)
	&=&
	0
\,.
\label{7VI22.1c}
\end{eqnarray}
Note that  $j^\beta \equiv 0$ for optical fibres, since the dielectric excludes currents in both regions.

Interface conditions can then be derived by requiring that no uncompensated Dirac-$\delta$ distributions appear in \eqref{7VI22.1a}--\eqref{7VI22.1c}. We additionally require that  $F_{\alpha \beta}$ has no $\delta$-distributions, so that the corresponding energy-momentum tensor is well-defined.
Note that $\gamma^{00} = - n(r)^2$ is explicitly discontinuous across the interface and thus $\p_r \gamma^{00} \propto \delta ( r - a)$.

First, the requirement that $F_{\alpha \beta}$ contain no Dirac-$\delta$ implies $\{A_0, A_\theta, A_z\} \in C( \R^4)$.
From this and \eqref{7VI22.1c} it follows that $A_r$ has to be continuous as well and further
\begin{equation}
	 r^{-1} \p_r (r A_r) + \t{ \m }{^0^0} \p_0 A_0 \in C(\R^4)
\,.
\label{7VI22.3}
\end{equation}

Next, examining \eqref{7VI22.1a} yields

\begin{eqnarray}
	r^{-1}\p_\alpha \left( r \t{ G }{^\alpha^0} \right) - \redmusquared A_0 \t{ \g }{^0^0}
	&=&
	0
	\qquad \Longrightarrow \qquad
	G^{r0} \in C(\R^4)
\,,
\label{7VI22.4a}
\\
	r^{-1}\p_\alpha \left( r \t{ G }{^\alpha^\theta} \right) - \redmusquared A_\theta \t{ \g }{^\theta^\theta}
	&=&
	0
	\qquad \Longrightarrow \qquad
	\p_r A_\theta \in C(\R^4)
\,,
\label{7VI22.4b}
\\
	r^{-1}\p_\alpha \left( r \t{ G }{^\alpha^z} \right) - \redmusquared A_z \t{ \g }{^z^z}
	&=&
	0
	\qquad \Longrightarrow \qquad
	\p_r A_z \in C(\R^4)
\,.
\label{7VI22.4c}
\end{eqnarray}

The case $\beta = r$ does not lead to additional constraints, since $G^{\alpha \beta}$ contains no $\delta$-terms and by anti-symmetry no radial derivative acts on it.

Finally, \eqref{7VI22.1b} does not contain Dirac-$\delta$ by continuity of the potential $A_\alpha$.

In conclusion, the following conditions must be satisfied at the interface:
\begin{eqnarray}
	A_\alpha
	&\in&
	C(\R^4)
\,,
\label{7VI22.6a}
\\
	G^{r0}
	&\in&
	C(\R^4)
\,,
\label{7VI22.6b}
\\
	\p_r A_\theta
	&\in&
	C(\R^4)
\,,
\label{7VI22.6c}
\\
	\p_r A_z
	&\in&
	C(\R^4)
\,,
\label{7VI22.6d}
\\
	r^{-1} \p_r (r A_r) + \t{ \m }{^0^0} \p_0 A_0 &\in& C(\R^4)
\,.
\label{7VI22.6f}
\end{eqnarray}

\subsection{ Gordon Mass Term }

Treating first the case of a Gordon mass term $\m = \gamma$, we observe that the second term in equation \eqref{eq:wave} vanishes by constraint  \eqref{eq:constraint}.
Thus, we simply have
\begin{equation}
	\Box_\g A_\beta - \redmusquared A_\beta = 0
\,,
\label{16I23.1}
\end{equation}
in regions of constant $n$.

Considerations for the optical fibre are now a straightforward extension of the Maxwell case (compare~\cite{Love83}).
A Fourier decomposition in cylindrical coordinates $ ( t , r , \theta , z )$ of the form
\begin{equation}
	A_\alpha = a_\alpha(r) e^{i (\beta z + m \theta - \omega t)}\,,
\label{17III22.1}
\end{equation}
yields solutions as Bessel functions of first and second kind
\begin{align}
	a_t^\text{core} &= c_t^\text{core} J_m \left(U \tfrac{r}{a} \right)\,,
\label{10III22.13a}
		&
	a_t^\text{clad} &= c_t^\text{clad} K_m \left(W \tfrac{r}{a} \right)\,,
		\\
	a_+^\text{core} &= c_+^\text{core} J_{m+1} \left(U \tfrac{r}{a} \right)\,,
\label{10III22.13b}
		&
	a_+^\text{clad} &= c_+^\text{clad} K_{m+1} \left(W \tfrac{r}{a} \right)\,,
		\\
	a_-^\text{core} &= c_-^\text{core} J_{m-1} \left(U \tfrac{r}{a} \right)\,,
\label{10III22.13c}
		&
	a_-^\text{clad} &= c_-^\text{clad} K_{m-1} \left(W \tfrac{r}{a} \right)\,,
		\\
	a_z^\text{core} &= c_z^\text{core} J_m \left(U \tfrac{r}{a} \right)\,,
\label{10III22.13d}
		&
	a_z^\text{clad} &= c_z^\text{clad} K_m \left(W \tfrac{r}{a} \right)\,,
\end{align}
where we defined an adapted basis via
\begin{equation}
	a_\pm = \frac{1}{\sqrt 2} \left( a_r \pm \frac{i}{r} a_\theta \right)
	\,.
\end{equation}
Further, the $c$'s are constants and
\begin{equation}
 \label{9XII22.1}
  U = \sqrt{ a^2 \left( n_1^2 \omega^2 - \beta^2 - \redmusquared  \right) }
   \,,
    \qquad
    W = \sqrt{ - a^2 \left( n_2^2 \omega^2 - \beta^2 - \redmusquared  \right)}
     \,.
\end{equation}

The eight unknown constants can be determined by solving the system of interface conditions \eqref{7VI22.6a}-\eqref{7VI22.6f}. Non-trivial solutions thus correspond to the roots of the associated determinant
\begin{equation}
\left|
\left(
{\small
\begin{array}{cccccccc}
 1 & 0 & 0 & 0 & -1 & 0 & 0 & 0 \\
 0 & \mathcal{J}-\frac{m}{U^2} & -\mathcal{J}-\frac{m}{U^2} & 0 & 0 & \frac{m}{W^2}-\mathcal{K} &
   -\mathcal{K}-\frac{m}{W^2} & 0 \\
 0 & \frac{m}{U^2}-\mathcal{J} & -\mathcal{J}-\frac{m}{U^2} & 0 & 0 & \mathcal{K}-\frac{m}{W^2} &
   -\mathcal{K}-\frac{m}{W^2} & 0 \\
 0 & 0 & 0 & 1 & 0 & 0 & 0 & -1 \\
 -\mathcal{J} U^2 & \frac{m}{U^2}-\mathcal{J} & \mathcal{J}+\frac{m}{U^2} & 0 & \frac{n_1^2}{n_2^2}
   \mathcal{K} W^2 & \frac{n_1^2}{n_2^2} \left(\mathcal{K}-\frac{m}{W^2}\right) & \frac{n_1^2}{n_2^2}
   \left(\mathcal{K}+\frac{m}{W^2}\right) & 0 \\
 0 & 1 & 1 & 0 & 0 & 1 & -1 & 0 \\
 0 & 0 & 0 & \mathcal{J} U^2 & 0 & 0 & 0 & -\mathcal{K} W^2 \\
 n_1^2 a^2 \omega ^2 & 1 & -1 & \beta  a ^2 \omega  & -n_2^2 a ^2 \omega ^2 & 1 & 1 & -\beta 
   a^2 \omega  \\
\end{array}
}
\right)
\right|
 \,,
\end{equation}
introducing new variables
\begin{equation}
	\mathcal{J}=\frac{J_m'(U)}{U J_m(U)}\,, \qquad \mathcal{K}=\frac{ K_m'( W ) }{ W K_m( W ) }\,,
	\label{eq:JK}
\end{equation}
as well as
\begin{equation}
	\psi = m \frac{ ( U^2 + W^2 ) }{ U^2 W^2 } \frac{ \beta }{ \omega }\,.
\end{equation}

The non-trivial mode solutions correspond to solutions of
\begin{equation}
	(\mathcal J + \mathcal K) ( n_1^2 \mathcal J + n_2^2 \mathcal K) = \frac{\redmusquared  + \beta^2}{\beta^2} \psi^2 \,,
 \label{8IV22.1}
\end{equation}
where we recover the usual Maxwell expression in the limit $\redmu \to 0$.

Given $\omega$, $m$, $a$, $n_1$ and $n_2$, the determinant equation  with $\redmu=0$ has finitely many solutions $\beta_\kappa$, where $\kappa$ is the radial mode index. One expects that, generically the left-hand side, viewed as a function of $\beta$, will have a first order root at the desired value $\beta_\kappa$; and this can be numerically checked for values of parameters  of interest. In such cases a nearby solution of \eqref{8IV22.1} will also exist for all $\redmu$ sufficiently small.

\begin{figure}[t]
    \centering
    \includegraphics[width=0.9\textwidth]{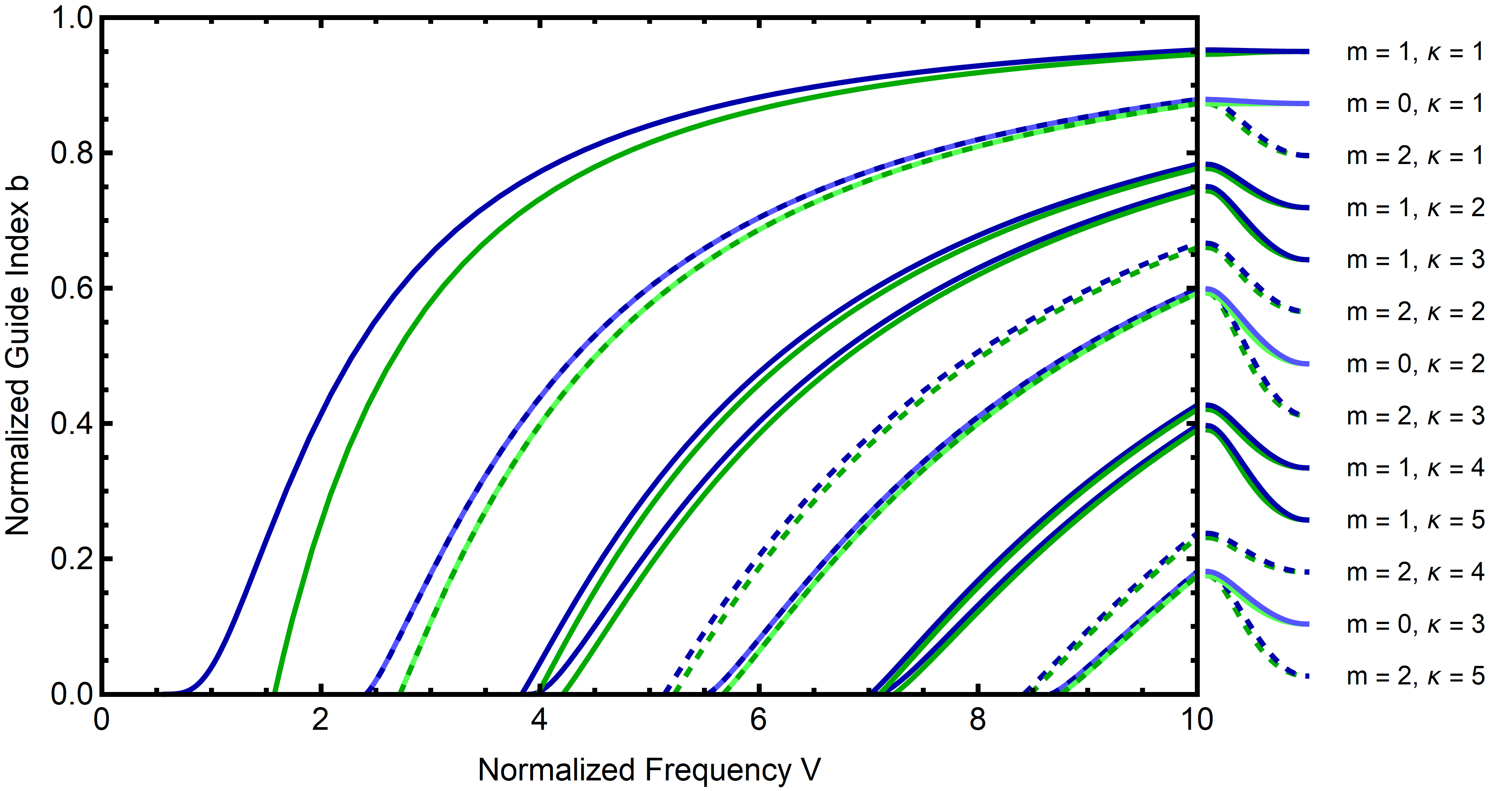}
    \caption{Mode spectrum with shades of blue for the Maxwell limit $\redmu = 0\,\mathrm{eV}$ and shades of green for the Gordon mass term with $\redmu = 1.5\,\mathrm{eV}$ for illustration purposes.
    The modes are characterized by their angular and radial mode index $m$ and $\kappa$ respectively. 
    Refractive indices are set to $n_1 = 1.4712$, $n_2 = 1.4659$ and the core radius to $ a = 1\,\mathrm{\upmu m}$.}
    \label{fig:modes}
\end{figure}

It is customary to use normalized frequency $V$ and normalized guide index $b$ defined as
\begin{equation}
	V= a \omega \sqrt{n_1^2 - n_2^2}\,,
	\qquad
	b = \frac{{\bar n}^2 - n_2^2}{n_1^2 - n_2^2}\,,
\end{equation}
where $\bar n = |\tfrac{\beta}{\omega}|$ is the effective refractive index. With these parameters we can visualize the mode spectrum and compare to ordinary Maxwell -- see Fig.~\ref{fig:modes}.

Further, explicitly calculating the corresponding nullspace, yields coefficients which also depend continuously upon $\redmu$:
\begin{align}
	c_t^\text{core} &= J_m^{-1} \left(U \right)\,,
\label{8IV22.3a}
		&
	c_t^\text{clad} &= K_m^{-1} \left(W \right)\,,
		\\
	c_+^\text{core} &= -\frac{ i a  \left(\beta n_1^2 \omega (\mathcal{J}+\mathcal{K})+(\redmusquared  + \beta^2) \psi \right) }{ \sqrt{2} \beta  U (\mathcal{J}+\mathcal{K}) J_m(U) }\,,
\label{8IV22.3b}
		&
	c_+^\text{clad} &= \frac{i a  \left(\beta n_2^2 \omega  (\mathcal{J}+\mathcal{K})+(\redmusquared +\beta^2) \psi \right)}{\sqrt{2} \beta  W (\mathcal{J}+\mathcal{K})
   K_m(W)}\,,
		\\
	c_-^\text{core} &= \frac{i a
   \left(\beta  n_1^2 \omega  (\mathcal{J}+\mathcal{K}) - (\redmusquared  + \beta^2) \psi \right)}{\sqrt{2} \beta  U (\mathcal{J}+\mathcal{K}) J_m(U)}\,,
\label{8IV22.3c}
		&
	c_-^\text{clad} &= \frac{i a \left(\beta  n_2^2 \omega  (\mathcal{J}+\mathcal{K}) - (\redmusquared  + \beta^2) \psi \right)}{\sqrt{2} \beta  W (\mathcal{J}+\mathcal{K}) K_m(W)}\,,
		\\
	c_z^\text{core} &= 0\,,
\label{8IV22.3d}
		&
	c_z^\text{clad} &= 0\,.
\end{align}
Taking $\redmu \to 0$ this solution is equivalent to the spectrum of physical solutions in the massless case for Lorenz gauge.

\subsection{ Minkowski Mass Term }

In the case of a Minkowski mass term $\m = \eta$ the divergence term in \eqref{eq:wave} does not vanish and we need to consider the wave equation
\begin{equation}
	\Box_\g A_\beta - (1 - n^2) \p _\beta  \p _0 A_0 - \redmusquared A_\sigma \t{ \eta }{^\sigma^\rho} \t{ \g }{_\rho_\beta} = 0
\,,
\label{16I23.2}
\end{equation}
in regions of constant $n$.

Using again the Fourier decomposition \eqref{17III22.1} we find
\begin{align}
	a_t^\text{core} &= c_t^\text{core} J_m \left(\tilde U \tfrac{r}{a} \right)\,,
\label{10III22.13a.2}
		&
	a_t^\text{clad} &= c_t^\text{clad} K_m \left(\tilde W \tfrac{r}{a} \right)\,,
		\\
	a_+^\text{core} &= c_+^\text{core} J_{m+1} \left(U \tfrac{r}{a} \right) + \tfrac{i \redsigma}{\sqrt{2}}\ladder_+ a_t^\text{core}\,,
\label{10III22.13b.2}
		&
	a_+^\text{clad} &= c_+^\text{clad} K_{m+1} \left(W \tfrac{r}{a} \right) + \tfrac{i \redtau }{ \sqrt{2}} \ladder_+  a_t^\text{clad} \,,
		\\
	a_-^\text{core} &= c_-^\text{core} J_{m-1} \left(U \tfrac{r}{a} \right) + \tfrac{i \redsigma}{\sqrt{2}} \ladder_- a_t^\text{core}\,,
\label{10III22.13c.2}
		&
	a_-^\text{clad} &= c_-^\text{clad} K_{m-1} \left(W \tfrac{r}{a} \right)+ \tfrac{i \redtau }{ \sqrt{2} (n_2^2} \ladder_- a_t^\text{clad}\,,
		\\
	a_z^\text{core} &= c_z^\text{core} J_m \left(U \tfrac{r}{a} \right)  - \beta \redsigma a_t^\text{core}\,,
\label{10III22.13d.2}
		&
	a_z^\text{clad} &= c_z^\text{clad} K_m \left(W \tfrac{r}{a} \right) - \beta \redtau  a_t^\text{clad}\,,
\end{align}
with $U$ and $W$ as in \eqref{9XII22.1},
\begin{equation}
 \label{16I23.3}
  \tilde U = \sqrt{ a^2 \left( \omega^2 - \beta^2 - n_1^{-2} \redmusquared  \right) }
   \,,
    \qquad
    \tilde W = \sqrt{ - a^2 \left( \omega^2 - \beta^2 - n_2^{-2} \redmusquared  \right)}
     \,,
\end{equation}
and
\begin{equation}
 \label{16I23.4}
  \redsigma =  \frac{\omega n_1^2}{(n_1^2 \omega^2 - \redmusquared)}
   \,,
    \qquad
    \redtau = \frac{\omega n_2^2}{(n_2^2 \omega^2 - \redmusquared)}
    \,,
    \qquad
    \ladder_\pm =  \p_r   \mp  \tfrac{m}{r}
     \,.
\end{equation}

Again the unknown constants must be determined by solving the interface conditions \eqref{7VI22.6a} - \eqref{7VI22.6f}, equivalent to the determinant condition
\begin{equation}
\left|\left(
{\small
\begin{array}{cccccccc}
 \tilde{U}^2+\beta ^2 a ^2 & 0 & 0 & 0 & \tilde{W}^2-\beta ^2 a ^2 & 0 & 0 & 0 \\
 \tilde{\mathcal{J}} \tilde{U}^2 & \mathcal{J}-\frac{m}{U^2} & \mathcal{J}+\frac{m}{U^2} & 0 & \tilde{\mathcal{K}} \tilde{W}^2 & \frac{m}{W^2}-\mathcal{K} &
   \mathcal{K}+\frac{m}{W^2} & 0 \\
 m & \frac{m}{U^2}-\mathcal{J} & \mathcal{J}+\frac{m}{U^2} & 0 & m & \mathcal{K}-\frac{m}{W^2} & \mathcal{K}+\frac{m}{W^2} & 0 \\
 1 & 0 & 0 & 1 & 1 & 0 & 0 & 1 \\
 \beta ^2 a ^2 \tilde{\mathcal{J}} \tilde{U}^2 & -\mathcal{J} U^2 & - \mathcal{J} U^2& 0 & \beta ^2 a ^2 \tilde{\mathcal{K}} \tilde{W}^2 &
   -\mathcal{K} W^2 & \mathcal{K} W^2 & 0 \\
 0 & 1 & -1 & 0 & 0 & 1 & 1 & 0 \\
 \tilde{\mathcal{J}} \tilde{U}^2 & 0 & 0 & \mathcal{J} U^2 & \tilde{\mathcal{K}} \tilde{W}^2 & 0 & 0 & \mathcal{K} W^2 \\
 0 & 1 & 1 & \beta ^2 a^2 & 0 & 1 & -1 & \beta ^2 a ^2 \\
\end{array}
}
\right)
\right|
= 0\,,
\end{equation}
where in addition to $\mathcal{J}$ and $\mathcal{K} $ as in \eqref{eq:JK}, we introduce
\begin{equation}
	\tilde{ \mathcal{J}}=\frac{J_m'( \tilde{U})}{ \tilde{U} J_m(\tilde{U})}\,, \qquad \tilde{\mathcal{K}}=\frac{ K_m'(\tilde  W ) }{ \tilde{W} K_m( \tilde W ) }\,.
	\label{eq:JK2}
\end{equation}
Due to the increased complexity we limit ourselves to the lowest order terms in the photon mass, for which the roots of the determinant are given by
\begin{align}
0 =  & (\mathcal{J} +\mathcal{K})   \big(\mathcal{J} n_1^2 +\mathcal{K} n_2^2\big) -\psi ^2
\\\notag
   & - \redmusquared \left( n_1^2 n_2^2 \omega ^2 \left(\omega ^2-\beta ^2\right) \left(\tilde{\mathcal{J}}+\tilde{\mathcal{K}}\right)\right)^{-1}
   \left(
     (\mathcal{J}+\mathcal{K})  \big(\mathcal{J} n_1^2  \mathscr{A} +\mathcal{K} n_2^2  \mathscr{B} + \mathcal{J} \mathcal{K} \mathscr{C} \big)
   -\psi ^2 \mathscr{D}
   \right)
   \\\notag
   &+ \mathscr{O}(\redmu^4)
    \,,
 \label{8IV22.1.2}
 \end{align}
 with
 \begin{align}
 \notag
 \mathscr{A} = & \beta ^2 \left(n_2^2 \left(\tilde{\mathcal{J}}+2 \tilde{\mathcal{K}}\right)+n_1^2 \tilde{\mathcal{J}}\right)-\left(n_1^2+2 n_2^2\right) \omega ^2
   \left(\tilde{\mathcal{J}}+\tilde{\mathcal{K}}\right)
   \,,
   \\\notag
    \mathscr{B} = & \beta ^2 \left(n_1^2 \left(2 \tilde{\mathcal{J}}+\tilde{\mathcal{K}}\right)+n_2^2 \tilde{\mathcal{K}}\right)-\left(2 n_1^2+n_2^2\right) \omega ^2
   \left(\tilde{\mathcal{J}}+\tilde{\mathcal{K}}\right)
   \,,
      \\\notag
    \mathscr{C} = & \beta^2 (n_1^2 - n_2^2 )^2 
   \,,
 \\\notag
 \mathscr{D}= &\mathcal{J} n_2^2 \beta^{-2} \left(\beta ^2-n_1^2 \omega ^2\right){}^2+\mathcal{K} n_1^2 \beta^{-2} \left(\beta ^2-n_2^2 \omega ^2\right){}^2
 + \left( \tilde{\mathcal{J}} n_1^2 + \tilde{\mathcal{K}} n_2^2 \right) (\beta^2-\omega^2) - \left(\tilde{\mathcal{J}} n_2^2 - \tilde{\mathcal{K}}  n_1^2\right) \omega^2
 \,.
\end{align}
In the limit $\redmu \to 0$ we recover the familiar Maxwell expression.

Although the determinant differs from the Gordon case \eqref{8IV22.1}, the roots are numerically in close agreement.
The kernel can be readily computed by a computer algebra system and leads to solutions which are small correction to the massless case.
However, due to the increased complexity there is no explicit form worth presenting. 

\section{Conclusions}

We construct solutions for Proca's equation in step-index optical fibres for two reasonable choices of mass tensor $\m = \{\gamma, \eta\}$, finding that the limit $\redmu \to 0$ agrees with the Maxwell description of a massless photon.
On grounds of naturalness we argue that the choice $\m = \gamma$ is preferable, since it leads to solutions in Gordon-Lorenz gauge and a significantly cleaner description.

Lastly, the no-go theorem for Proca in coaxial waveguides~\cite{CMS22} does not extend to the optical fibre case, which only sees $O(\redmu^2 a^2)$ corrections.

\section*{Acknowledgements}
Research supported in part by the Austrian Science Fund (FWF) Project P34274, as well as the Vienna University Research Platform TURIS. I am very grateful to Piotr Chru\'sciel and Thomas Mieling for many useful discussions.

\providecommand{\bysame}{\leavevmode\hbox to3em{\hrulefill}\thinspace}
\providecommand{\MR}{\relax\ifhmode\unskip\space\fi MR }
\providecommand{\MRhref}[2]{%
  \href{http://www.ams.org/mathscinet-getitem?mr=#1}{#2}
}
\providecommand{\href}[2]{#2}


\begin{thebibliography}{10}

\bibitem{AdelbergerDvaliGruzinov}
E.~{Adelberger}, G.~{Dvali}, and A.~{Gruzinov}, \emph{{Photon-Mass Bound
  Destroyed by Vortices}}, \prl\ \textbf{98} (2007), no.~1, 010402.

\bibitem{BONETTI2017203}
L.~{Bonetti}, L.R. {dos Santos Filho}, J.A. {Helay{\"e}l-Neto}, and A.D.A.M.
  {Spallicci}, \emph{{Effective photon mass by Super and Lorentz symmetry
  breaking}}, Physics Letters B \textbf{764} (2017), 203--206.

\bibitem{Goldhaber08}
A.~S. Goldhaber and M.~M. Nieto, \emph{{Photon and Graviton Mass Limits}}, Rev.
  Mod. Phys. \textbf{82} (2010), 939--979.

\bibitem{Gordon23}
W.~{Gordon}, \emph{{Zur Lichtfortpflanzung nach der Relativit{\"a}tstheorie}},
  Annalen der Physik \textbf{377} (1923), no.~22, 421--456.

\bibitem{Hilweg17}
C.~Hilweg, F.~Massa, D.~Martynov, N.~Mavalvala, P.~T. Chru\'sciel, and
  P.~Walther, \emph{{Gravitationally induced phase shift on a single photon}},
  New J. Phys. \textbf{19} (2017), no.~3, 033028.

\bibitem{Jackson98}
J.~D. Jackson, \emph{Classical electrodynamics}, Wiley, 1998.

\bibitem{Mieling22}
T.~B. Mieling, C.~Hilweg, and P.~Walther, \emph{{Measuring space-time curvature
  using maximally path-entangled quantum states}}, Phys. Rev. A \textbf{106}
  (2022), no.~3, L031701.

\bibitem{Proca36}
A.~Proca, \emph{{Sur la theorie ondulatoire des electrons positifs et
  negatifs}}, J. Phys. Radium \textbf{7} (1936), 347--353.

\bibitem{Love83}
A.~W. Snyder and J.~Love, \emph{Optical waveguide theory}, Springer, 1983.

\bibitem{CMS22}
F.~Steininger, T.~Mieling, and P.~T. Chruściel, \emph{No proca photons}, 2022.

\bibitem{ParticleDataGroup:2022pth}
R.~L. Workman et~al., \emph{{Review of Particle Physics}}, PTEP \textbf{2022}
  (2022), 083C01.

\end{thebibliography}
\end{document}